\documentclass[12pt]{elsarticle}
\usepackage{graphicx}
\usepackage{amsmath}
\usepackage{amssymb}
\usepackage{gensymb}
\usepackage{xcolor}
\usepackage[nooneline,tight]{subfigure}
\DeclareMathOperator{\ch}{ch}
\DeclareMathOperator{\sh}{sh}
\DeclareMathOperator{\tnh}{th}

\mathchardef\mhyphen="2D

\newcommand{\Ham}{\mathcal{H}}     
 
\begin{document}


\title
{Exact solutions to plaquette Ising models with free and periodic boundaries}

\author[itp]{Marco Mueller}
\ead{Marco.Mueller@itp.uni-leipzig.de}
\author[hwu]{Desmond A. Johnston}
\ead{D.A.Johnston@hw.ac.uk}
\author[itp]{Wolfhard Janke}
\ead{Wolfhard.Janke@itp.uni-leipzig.de}
\address[itp]{Institut f\"ur Theoretische Physik, Universit\"at Leipzig,\\ Postfach 100\,920, D-04009 Leipzig, Germany}
\address[hwu]{Department\ of Mathematics and the Maxwell Institute for Mathematical Sciences, Heriot-Watt University, Riccarton, Edinburgh, EH14 4AS, Scotland}


\begin{abstract}
An anisotropic limit of the $3d$ plaquette Ising model, in which the plaquette
couplings in one direction were set to zero, was solved  for free boundary
conditions by Suzuki (Phys.~Rev.~Lett. {\bf 28} (1972) 507), who later dubbed
it the fuki-nuke, or ``no-ceiling'', model.  Defining new spin variables as the
product of nearest-neighbour spins transforms the Hamiltonian  into that of a
stack of (standard) $2d$ Ising models and reveals the planar nature of the
magnetic order, which is also present in the fully isotropic $3d$ plaquette
model.  More recently, the solution of the fuki-nuke model was  discussed for
periodic boundary conditions, which require  a different approach to defining
the product spin transformation, by Castelnovo et~al.  (Phys. Rev. B {\bf 81}
(2010) 184303).
	
We clarify the exact relation between partition functions with free and
periodic boundary conditions expressed in terms of original and product spin
variables for the $2d$ plaquette and $3d$ fuki-nuke models, noting that the
differences are already present in the $1d$ Ising model.  In addition, we solve
the $2d$ plaquette Ising model with helical boundary conditions. The various
exactly solved examples illustrate how correlations can be induced in finite
systems as a consequence of the choice of boundary conditions.

\end{abstract}


 
\maketitle


\section{Introduction}

The strongly anisotropic limit of a purely plaquette Ising Hamiltonian  on a
$3d$ cubic lattice
\begin{equation}
\label{eq:ham:gonikappa0}
\Ham= - J \sum_{\Box}  \sigma \sigma \sigma \sigma \;,
\end{equation}
where we denote the product of the spins sited at vertices around a plaquette
by $\Box$ and in which the plaquette coupling $J$ in one direction is set to
zero, may be solved exactly~\cite{suzuki_old,castelnovo}. A variable
transformation in which a product of nearest-neighbour spins in the direction
perpendicular to the \mbox{non-contributing} plaquettes is made, $\hat\tau_i =
\sigma_{i-1} \sigma_i$, reveals that the model (later dubbed fuki-nuke by
Hashizume~and~Suzuki~\cite{suzuki1})  is \mbox{non-trivially} equivalent to a
stack of standard $2d$ Ising models with nearest-neighbour pair interactions in
each plane. 

The nature of the order in the fuki-nuke model is rather unusual since the
$\tau$-spins may magnetize independently in each $2d$ Ising plane. In terms of
the original $\sigma$ spins this order is encoded in nearest-neighbour
correlators perpendicular to the direction in which the plaquette coupling is
zero.  An isotropic version of this planar order exists for the isotropic
plaquette Hamiltonian~\cite{suzuki1, us_fukinuke}. The isotropic model in
Eq.~(\ref{eq:ham:gonikappa0}) has a strong first-order phase
transition~\cite{firstorder} with several interesting properties itself. It
displays non-standard finite-size scaling because of its exponentially
degenerate low-temperature phase~\cite{us_goni} and it also has  glassy
characteristics~\cite{goni_glassy}, in spite of the absence of any quenched
disorder.  It can also be thought of as a particular limit of a family of
gonihedric \cite{savvidy1, savvidy2} Ising models containing
nearest-neighbour, next-to-nearest-neighbour and plaquette interactions tuned
to remove the bare area contribution of (geometric) spin clusters.

Suzuki's original solution of the fuki-nuke model employed free boundary
conditions~\cite{suzuki_old}, whereas  periodic boundary conditions,
as often used in numerical simulations, were considered in
\cite{castelnovo}. Although the treatment of the product variable
transformation $\hat\tau_i = \sigma_{i-1} \sigma_{i}$ in the two cases can,
loosely, be argued to be identical in the thermodynamic limit as a post-hoc
justification for ignoring any 
subtleties,
it is possible to treat the
variable transformation for both free and periodic boundary conditions in  the
fuki-nuke model exactly and we shall do so here.  Since such differences
arising from boundary conditions may impact the finite-size scaling properties
in simulations, careful consideration of both cases is worthwhile.
 
Interestingly, the  differences  arising in using the product variable
transformation for free and periodic boundary conditions are already present
for the  nearest-neighbour $1d$ Ising model.  While the product spin
transformation has been widely used to obtain the solution of the $1d$ Ising
model for {\it free} boundary conditions, the discussion of \emph{periodic}
boundaries where constraints must be imposed on the allowed spin configurations
is less well known%
\footnote{
  The only previous discussion we have been made aware of is contained in
  lecture notes by Turban \cite{loic}.
} 
and the differences between the two are pointed out in preparation of similar
calculations in the $2d$ plaquette and $3d$ fuki-nuke models in 
Sec.~\ref{sec:onedising},
which also serves as a reference when expressing the solutions to the more
complicated models in terms of $1d$ Ising partition functions.

In Sec.~\ref{sec:fss2d}, we investigate the finite-size behaviour of the $2d$
plaquette Ising model, which appears as a different anisotropic limit of the
$3d$ plaquette Ising model using the product spin approach.  The exact solution
of the  $2d$ plaquette Ising model illustrates clearly how non-trivial
correlations can enter finite systems as a consequence of the choice of the
boundary conditions

In Sec.~\ref{sec:anisofuki}, we discuss the $3d$ anisotropic, fuki-nuke model
itself, following closely the $1d$ Ising template, since the technical issues
are
similar.
As with the $2d$ plaquette model, periodic boundary conditions
are found to induce correlations in the finite-size  fuki-nuke model  in
comparison to the (simpler) case of free boundaries.   Along the way we present
an exact numerical enumeration of the partition function, confirming the
equality of the expressions in terms of the original, $\sigma$, and product,
$\tau$, spins in the case of the fuki-nuke model. Finally,
Sec.~\ref{sec:conclusions} contains our conclusions.

\section{One Dimension: The Standard $1d$ Ising Model}
\label{sec:onedising}

The $1d$ Ising model provides perhaps {\it the} standard pedagogical
example of an exactly solvable model in statistical mechanics, albeit one
without a phase transition at finite temperature, as Ising himself
discovered~\cite{Ising} to his disappointment. It is often discussed using
periodic boundary conditions and a transfer matrix approach, since this allows
a straightforward solution, even in non-zero external field.  With a view to
the solution of  the fuki-nuke model we consider the model in zero external
field and take a different approach, in effect changing the variables in
the partition function so that it takes a factorized form and may be evaluated
trivially. The steps required to do this differ for the case of free and
periodic boundary conditions and we deal with each separately.   
 
\subsection{Free Boundary Conditions}

If we consider the standard nearest-neighbour Ising Hamiltonian with spins 
$\sigma_i=\pm 1$ on a linear chain of length $L$ in one dimension
\begin{equation}
H = -  \sum_{i=1}^{L-1} \sigma_i \sigma_{i+1}
\end{equation}
with free boundary conditions, then the partition function
\begin{equation}
    Z_{1d,\; {\rm free}} = \sum_{ \{\sigma \}}  \exp \left( \beta \sum_{i=1}^{L-1} \sigma_i \sigma_{i+1} \right)
\end{equation}
may be evaluated by defining the variable transformation
\begin{equation}
\{ \sigma_1, \sigma_2, \ldots \sigma_{L} \} \rightarrow 
\{ \tau_1,  \tau_2, \ldots \tau_L  \}\;,
\end{equation}
where $\tau_1 = \sigma_1 \sigma_2, \, \tau_2 = \sigma_2 \sigma_3,\, \ldots,
\tau_{L-1} = \sigma_{L-1} \sigma_L$. Setting $\tau_L = \sigma_L$ the mapping
$\{\sigma\} \rightarrow \{\tau\}$ with an inverse relation of the form
$\sigma_{i} = \tau_{L} \, \tau_{L-1} \, \tau_{L-2} \cdots \tau_{i}$ is
one-to-one.  This allows us to write $Z$ in factorized form as
\begin{equation}
Z_{1d,\; {\rm free}} = \sum_{\{\tau\}}\exp \left(  \beta \sum_{i=1}^{L-1} \tau_i \right) 
\end{equation}
which may then trivially be evaluated to give
\begin{equation}
\label{1DF}
  Z_{1d,\; {\rm free}} = 2 \prod_{i=1}^{L-1} \sum_{\tau_i = \pm 1} \exp \left( \beta \tau_i \right) = 2 ( 2 \ch(\beta))^{L-1}
\end{equation}
where the initial factor of two comes from the sum over $\tau_L = \sigma_L$
which does not appear in the exponent.  We highlight two features of this
calculation, which also appear when the transformation is applied to the
fuki-nuke model with free boundaries:
\begin{itemize}
	\item{} The last spin, $\sigma_L$, remains untransformed,
  \item{} summing over this gives a factor of $2$ in $Z_{1d,\;{\rm free}}$. 
\end{itemize}

\subsection{Periodic Boundary Conditions}

When periodic boundary conditions are imposed, we  map $L$ $\sigma$'s to $L$
$\tau$'s without requiring the condition $\tau_L = \sigma_L$ of the
free boundary conditions.  Since every configuration of $\tau$'s can now be
made up from two configurations of $\sigma$'s, this should be taken into
account when relating the partition functions expressed in terms of $\sigma$ or
$\tau$.  Explicitly, the transformations are now given by $\tau_1 = \sigma_1
\sigma_2, \, \tau_2 = \sigma_2 \sigma_3,\, \ldots,  \tau_L = \sigma_{L}
\sigma_{L+1} = \sigma_{L} \sigma_{1}$, with an inverse relation of the form
$\sigma_{i} =\sigma_{1}\times \tau_{1} \, \tau_{2} \, \tau_{3} \cdots
\tau_{i-1}$, and a direct consequence of the periodic boundary conditions is
that the constraint
\begin{equation}
    \prod_{i=1}^L \tau_i = \prod_{i=1}^L \sigma_i^2 = 1
\end{equation} 
must be imposed on the $\tau$-variables. This can be implemented in the
partition  function as 
\begin{equation}
\label{tauconstraint}
    Z_{1d,\;{\rm periodic}} = 2 \sum_{\{ \tau   \}}\exp \left(  \beta \sum_{i=1}^{L} \tau_i
\right) \delta \left( \prod_{i=1}^L \tau_i, 1 \right) \;,
\end{equation}
where the requisite factor of two takes account of the two-to-one
$\sigma$-to-$\tau$-mapping.  Since $\prod_{i=1}^L\tau_i=\pm1$, it is possible
to rewrite the Kronecker-$\delta$ function appearing in
Eq.~(\ref{tauconstraint}) as
\begin{equation}
Z_{1d,\;{\rm periodic}} = \sum_{\{ \tau   \}}\exp \left(  \beta \sum_{i=1}^{L} \tau_i
 \right) \left( 1 + \prod_{i=1}^L \tau_i  \right) \, 
\end{equation}
which subsumes the factor of two.  The partition function written in this form
may now be straightforwardly evaluated as the sum of two factorized terms,
\begin{eqnarray}
\label{proddelta}
Z_{1d,\;{\rm periodic}} &=& \left[ \prod_{i=1}^L \sum_{\tau_i = \pm 1} \exp \left( \beta \tau_i \right) + \prod_{i=1}^L \sum_{\tau_i = \pm 1} \tau_i \exp \left( \beta \tau_i \right) \right] \nonumber \\
&=& 2^L \left[ \ch(\beta)^L + \sh(\beta)^L \right]\nonumber\\
&=& 2^L \ch(\beta)^L \left[ 1 + \tnh(\beta)^L \right]\;.
\end{eqnarray}
The standard result for periodic boundary conditions, familiar from the
transfer matrix calculation and numerous other approaches, is hence recovered.
In the case of periodic boundary conditions we can see that:
\begin{itemize}
	\item{} The last spin, $\sigma_L$, \emph{is} included in the transformation,
  \item{} an additional factor of two appears in order to ensure
    the equivalence of the  $\sigma$- and $\tau$-representations of the
    partition function,
  \item{} a constraint must be imposed on the product of all the
    $\tau$-variables resulting in two terms in the partition function,
    corresponding to an additional correlation by comparison with free boundary
    conditions.
\end{itemize}
The factor of two thus appears for different reasons in the
$\tau$-representation of the partition function in the free boundary case
(summing over the last spin) and the periodic boundary case (a two-to-one
mapping between $\sigma$'s and $\tau$'s). 

\section{Two Dimensions: The $2d$ Gonihedric Ising Model}
\label{sec:fss2d}

Consider the {\it anisotropic} version of the Hamiltonian in
Eq.~(\ref{eq:ham:gonikappa0}),
\begin{eqnarray}
 \label{aniso}
H_{\rm aniso}  (\{ \sigma \} ) &=& - J_x  \sum_{x=1}^{L_x}
 \sum\limits_{y=1}^{L_y}\sum\limits_{z=1}^{L_z} \sigma_{x,y,z} \sigma_{x,y+1,z}\sigma_{x,y+1,z+1} \sigma_{x,y,z+1} \nonumber \\
&{}&  - J_y \sum_{x=1}^{L_x} \sum\limits_{y=1}^{L_y}\sum\limits_{z=1}^{L_z}  \sigma_{x,y,z} \sigma_{x+1,y,z}\sigma_{x+1,y,z+1} \sigma_{x,y,z+1}  \\
&{}& - J_z \sum_{x=1}^{L_x} \sum\limits_{y=1}^{L_y}\sum\limits_{z=1}^{L_z}  \sigma_{x,y,z} \sigma_{x+1,y,z}\sigma_{x+1,y+1,z} \sigma_{x,y+1,z} \nonumber
 \;,
\end{eqnarray}
where we have now  indicated each site and directional sum explicitly. If we
now set the coupling of the  vertical plaquettes to zero, the different
horizontal layers decouple trivially and the Hamiltonians of the individual
layers are those of the two-dimensional plaquette (gonihedric) \cite{savvidy1,
savvidy2} model,
\begin{equation}
H_{\rm aniso}^{J_x=J_y=0} (\{ \sigma \} ) = -
J_z \sum\limits_{z=1}^{L_z} \left[\;\sum_{2d\;\Box} \sigma\sigma\sigma\sigma \right] \; .
\end{equation}
Taking $J_z=1$ for simplicity, the partition function is  given by the
product of $L_z$ decoupled layers,
\begin{equation}
  Z_{\rm aniso}^{J_x = J_y = 0} = \sum_{\{ \sigma \}} \exp \left(- \beta H_{\rm aniso}^{J_x=J_y=0} (\{ \sigma \} ) \right) = (Z_{2d,\;{\rm gonihedric}})^{L_z} \;,
\end{equation}
each of which  is a $2d$ plaquette  model. The partition function for the $2d$
plaquette  model may also be evaluated exactly using the spin-bond
($\sigma$-$\tau$)-transformation for both free and periodic boundary conditions
in the direction of the transformation, which we shall take in in the following
along the vertical $y$-axis.

\subsection{Free Boundary Conditions in $y$-Direction}
\label{sec:fss2dfree}

On a rectangular $L_x\times L_y$ lattice with free boundaries 
in the $y$-direction, the $\sigma$-$\tau$-transformation used in the $1d$
Ising model can still be applied in $y$-direction by defining $\tau_{x,y} =
\sigma_{x,y}\sigma_{x,y+1}$, with the condition $\tau_{x,L_y} = \sigma_{x,L_y}$
and the inverse relation $\sigma_{x,y} = \tau_{x,L_y}\tau_{x,L_y-1}\cdots\tau_{x,y}$.
Assuming free boundaries in $x$-direction, too, the partition function reads
\begin{eqnarray}
  \label{2dgoni_free}
	Z_{2d,\;{\rm gonihedric},\;{\rm free}} 
	&=& \sum_{\{\sigma\}} \exp\left(\beta\sum\limits_{x=1}^{L_x-1}\sum\limits_{y=1}^{L_y-1}\sigma_{x,y}\sigma_{x,y+1}\sigma_{x+1,y}\sigma_{x+1,y+1}\right)\nonumber\\
	&=& \sum_{\{\tau\}} \exp\left(\beta\sum\limits_{x=1}^{L_x-1}\sum\limits_{y=1}^{L_y-1} \tau_{x,y}\tau_{x+1,y}\right)\nonumber\\
	&=& 2^{L_x} (Z_{1d,\;{\rm Ising}})^{L_y-1}\;,
\end{eqnarray}
where the factor $2^{L_x}$ in the last line comes from the $L_x$ sums over
$\tau_{x,L_y}=\sigma_{x,L_x}=\pm1$ which do not appear in the exponent, similar
to the $1d$ Ising case. Products of the partition function of the $1d$ Ising
model appear due to the decoupling in $\tau$-spins in the $y$-direction.
The solution of the free $1d$ Ising model from Eq.~(\ref{1DF}) simplifies this
expression to
\begin{eqnarray}
    \label{eq:2dgonihedric_free}
    Z_{2d,\;{\rm gonihedric},\;{\rm free,\;free}} 
	&=& 2^{L_x L_y} \ch\left(\beta\right)^{(L_x-1)(L_y-1)}\;.
\end{eqnarray}
With periodic boundary conditions in $x$-direction, the partition function
$Z_{1d,\;{\rm Ising}}$ in (\ref{2dgoni_free}) is the solution (\ref{proddelta})
of the periodic case, so that the explicit expression looks slightly more
complicated,
\begin{eqnarray}
  Z_{2d,\;{\rm gonihedric},\; {\rm periodic,\; free}} = 2^{L_x L_y} \ch\left(\beta\right)^{L_x (L_y-1)}\left(1+\tnh\left(\beta\right)^{L_x}\right)^{L_y-1}\nonumber\\
  \qquad = 2^{L_x L_y} \ch\left(\beta\right)^{L_x (L_y-1)}\sum_{h=0}^{L_y-1}{L_y-1\choose h}\tnh(\beta)^{L_x h}\;.
    \label{eq:2dgonihedric_mixed}
\end{eqnarray}
The expansion in the last line gives binomials of $\tnh(\beta)$, which also 
appear below when periodic boundaries in both directions are considered.

\subsection{Periodic Boundary Conditions in $y$-Direction}

To simplify the combinatorics involved when solving the model with periodic
boundary conditions, we  employ a dimer representation that allows us to
straightforwardly take into account the constraints that arise with periodic
boundaries. This diagrammatic approach appears naturally in the
high-temperature representation as a way of representing valid configurations
graphically.

If we take periodic boundary conditions in $y$-direction, i.e.,
$\sigma_{L_x+1,y} = \sigma_{1,y}$ and $\sigma_{x,L_y+1} = \sigma_{x,1}$, the
transformation $\tau_{x,y} = \sigma_{x,y}\sigma_{x,y+1}$
imposes the $L_x$ constraints $\prod_{y}\tau_{x,y} = 1$ and leads to an inverse
relation of the form 
\begin{equation}
\sigma_{x,y} =\sigma_{x,1}\times \tau_{x,1} \, \tau_{x,2} \, \tau_{x,3} \cdots \tau_{x,y-1}\; .
\end{equation}
This allows the partition function to be expressed in terms of the new
$\tau$-variables as
\begin{eqnarray}
  \label{Z_2d_goni_sigma_to_tau}
  Z_{2d,\;{\rm gonihedric},\;{\rm periodic}} 
  = \sum\limits_{\{\sigma\}}
    \exp\left(
    \beta\sum_{x=1}^{L_x}\sum_{y=1}^{L_y} 
        \sigma_{x,y}\sigma_{x,y+1}\sigma_{x+1,y}\sigma_{x+1,y+1}
    \right)\nonumber\\
  \qquad= 2^{L_x}\sum\limits_{\{\tau\}}
    \exp\left(
    \beta\sum_{x=1}^{L_x}\sum_{y=1}^{L_y} 
        \tau_{x,y}\tau_{x+1,y}
        \right)\prod_{x=1}^{L_x}\delta\left(\prod_{y=1}^{L_y}\tau_{x,y},1\right)\;,
\end{eqnarray}
where the prefactor of $2^{L_x}$ again accounts for the two-to-one
$\sigma$-to-$\tau$-mapping. The notation in
Eq.~(\ref{Z_2d_goni_sigma_to_tau}) assumes periodic boundary conditions also in
$x$-direction, although this is not essential for what follows (for free boundary
conditions we would have the replacement, $\sum_{x=1}^{L_x} \rightarrow
\sum_{x=1}^{L_x-1}$).

This can be rewritten in the high-temperature representation as an expression
which looks similar to the starting point of the combinatorial solution of the
standard $2d$ Ising model~\cite{combinatorial2dIsing},
\begin{flalign}
  \label{HighT}
 \qquad  Z_{2d,\;{\rm gonihedric},\;{\rm periodic}} &=&& \\ 
  \qquad 2^{L_x} \ch(\beta)^{L_x L_y} \sum\limits_{\{\tau\}} &
     \left[\,\prod_{y=1}^{L_y}\prod_{x=1}^{L_x}
     \left(1 + \tnh\left(\beta\right)\tau_{x,y}\tau_{x+1,y}\right)\right]\prod_{x=1}^{L_x}\delta\left(\prod_{y=1}^{L_y}\tau_{x,y},1\right).&&\nonumber
\end{flalign}
Here, however, we are saved from  the combinatorial complications of counting
loops because the spins only couple in the $x$ (horizontal) direction in our
case.  Graphically the factors of $\tnh(\beta) \tau_{x,y}\tau_{x+1,y}$, which
appear when expanding the product in Eq.~(\ref{HighT}), are represented as
horizontal dimers.  This amounts to the diagrammatical solution of the $1d$
Ising model using the high-temperature representation, up to subtle
complications due to the $\delta$-constraints discussed further below.

Let us first verify that, within this diagrammatic approach, the results of the
preceding subsection in
Eqs.~(\ref{eq:2dgonihedric_free})~and~(\ref{eq:2dgonihedric_mixed}) are
immediately recovered: For the case with free boundaries in both directions,
the $\delta$-constraints (and also the associated $2^{L_x}$ prefactor) are
absent, so $1,\dots,(L_x-1)\times(L_y-1)$ dimers cannot be arranged without any
dangling ends, since summing over the spins on the free dimer ends would give a
zero contribution to the partition function. This leaves the empty lattice as
the only contributing dimer configuration, giving the $2^{L_x L_y}$ factor in
Eq.~(\ref{eq:2dgonihedric_free}) from the then trivial summations over the
$\tau$-spins.  
In the other case with free boundaries in $x$ (horizontal) direction, and
periodic boundary conditions in $y$-direction, the direction in which the
$\sigma$-$\tau$-transformation is carried out, the $L_x$ $\delta$-constraints
couple the spins non-locally so that complete columns of dimers contribute,
too. There are $L_x -1 \choose v$ possible ways of choosing $v$ such columns,
each one carrying a weight of $\tnh(\beta)^{L_y v}$. Summing over all possible
numbers for $v$, the symmetric counterpart of Eq.~(\ref{eq:2dgonihedric_mixed})
is recovered, with $L_x$ and $L_y$ swapped (since here the
$\sigma$-$\tau$-transformation was carried out in the other direction). The
prefactor $2^{L_x L_y} = 2^{L_x}2^{L_x(L_y-1)}$ is the product of the factor
$2^{L_x}$ in Eq.~(\ref{HighT}) and the weight of $2^{L_x(L_y-1)}$ for each
diagram of dimers, which takes care of proper summation over all
$\tau$-configurations.  Here, for each of the $L_x$ spin columns (not to be
confused with the $L_x-1$ dimer columns), the $L_y$ summations over
$\tau_{x,y}$  give a trivial factor of $2$, except for one summation (say, the
first) which gives only 1 due to summing over the $\delta$-constraint.

\begin{figure}
	\centering
  \includegraphics{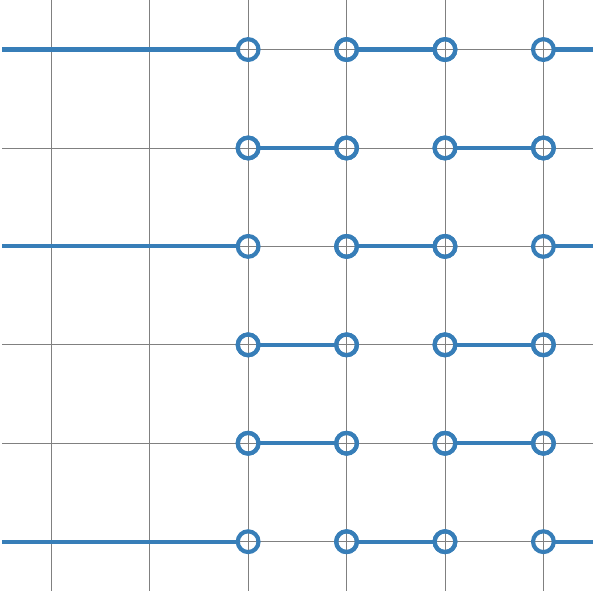}
  \caption{A contributing dimer configuration in the $2d$ gonihedric case with
    periodic boundary conditions in both directions for $L_x=L_y=6$, combining
    both $\tnh(\beta)^4$ and $\tnh(\beta)^2$ terms.}
  \label{Dimers}
\end{figure}

After these checks, we are ready to consider the doubly periodic case (i.e.,
the torus topology), where periodic boundary conditions are assumed in both
$x$- and $y$-direction and the graphical representation is slightly more
complicated.  Here not only empty but also completely filled rows (``closed''
by the periodic boundary conditions in $x$-direction) of dimers would normally
contribute. However, due to the $\delta$-constraints, gaps in the otherwise
filled rows of dimers may also be present. As a consequence, both a horizontal
configuration of dimers and its ``dual'', where shaded and unshaded bonds are
swapped, may appear.  In Fig.~\ref{Dimers} a contributing configuration to the
$L_x=L_y=6$ partition function is shown, where the dimers giving $\tnh (\beta)$
factors are shown heavily shaded.  The two sorts of contributing horizontal
lines give either an $\tnh(\beta)^4$ or $\tnh(\beta)^2$ factor in this case.
In general on an $L_x \times L_y$ lattice there may be $v=0,...,L_x$ gaps in a
shaded line which may be chosen in ${L_x \choose v}$ ways and counting these
and their duals gives \label{sec:dimergap}
\begin{flalign}
 \label{dimers_periodic}
 \qquad Z_{2d,\;{\rm gonihedric},\;{\rm periodic}} &=&& \\
 \qquad
\left(\frac{1}{2}\right)   &2^{L_x L_y} \ch(\beta)^{L_x L_y} \sum\limits_{v=0}^{L_x}{L_x \choose v}
  \left(\tnh(\beta)^v + \tnh(\beta)^{L_x-v}\right)^{L_y}\,.&&\nonumber
\end{flalign}
The prefactor of $1/2$ takes care of the double-counting inherent in
the dimer description due to the $v \leftrightarrow L_x -v$ symmetry. The
diagram of Fig.~\ref{Dimers}, for instance,  appears in both the $v=2$ and
$v=4$ terms in the sum of  Eq.~(\ref{dimers_periodic}). The other prefactor,
$2^{L_x L_y} = 2^{L_x}2^{L_x(L_y-1)}$ results as above from the now
trivial summations over the $\tau$-spins, respecting the $L_x$
$\delta$-constraints (which kill one of the factors of 2 for each 
of the $L_x$ spin columns).

Expanding the product in Eq.~(\ref{dimers_periodic}) gives an alternative
representation of the partition function as a double sum, which was also found
by Espriu~and~Prats~\cite{espriu_prats} for the special case $L_x=L_y=L$ by
enumerating possible {\it plaquette} configurations. In this approach rows and
columns of plaquettes which can contribute to the partition function sum are
counted, keeping track of over-counting factors of $\tnh(\beta)$ in
intersecting rows and columns:
\begin{flalign}
	\label{dimers_periodic2}
 \qquad Z_{2d,\;{\rm gonihedric},\;{\rm periodic}} &=&&\\
  \qquad  \left( \frac{1}{2} \right) & 2^{L_x L_y}\ch(\beta)^{L_x
  L_y}\sum\limits_{v=0}^{L_x}\sum_{h=0}^{L_y} {L_x \choose v} {L_y
  \choose h}
  \tnh(\beta)^{vL_y + hL_x - 2vh}\; . &&\nonumber
\end{flalign}
In \mbox{\ref{helicalsolution}} we show how enumerating plaquette
configurations also allows the exact solution of the model with helical
boundary conditions as considered recently in a numerical Monte Carlo 
simulation study \cite{davatolhagh}.

In summary, although we have used the same transformation, the solution for the
model with periodic boundary conditions can be seen to be more involved than
the (almost) trivial free case in Sec.~\ref{sec:fss2dfree}. This is a
consequence of the constraints that implement the periodic boundary conditions,
which couple the different $1d$ layers and allow non-trivial $1d$
configurations to contribute to the partition function sum.  As we will now
see, this behaviour is repeated in the three-dimensional fuki-nuke model where
free boundary conditions lead to a partition function composed of uncoupled
$2d$ layers, whereas periodic boundaries give a much more complicated
structure.

\section{Three Dimensions: The Fuki-Nuke Model}
\label{sec:anisofuki}

\subsection{The Fuki-Nuke Model}

The fuki-nuke model~\cite{suzuki_old,suzuki1} is the $J_z=0$ limit of the
anisotropic $3d$ plaquette model defined in Eq.~(\ref{aniso}). In this case the
horizontal, ``ceiling'' plaquettes have zero coupling, which Hashizume and
Suzuki denoted the  fuki-nuke (``no-ceiling'' in Japanese)
model~\cite{suzuki1}. The anisotropic $3d$ plaquette Hamiltonian when $J_z=0$
is thus given by
\begin{eqnarray}
H_{\rm fuki\mhyphen nuke} (\{ \sigma \} ) &=& - J_x  \sum_{x=1}^{L} \sum\limits_{y=1}^{L}\sum\limits_{z=1}^{L_z} \sigma_{x,y,z} \sigma_{x,y+1,z}\sigma_{x,y+1,z+1} \sigma_{x,y,z+1} \nonumber \\
&{}&  - J_y \sum_{x=1}^{L} \sum\limits_{y=1}^{L}\sum\limits_{z=1}^{L_z}  \sigma_{x,y,z} \sigma_{x+1,y,z}\sigma_{x+1,y,z+1} \sigma_{x,y,z+1}\;,
\label{eq:fuki-nuke-sigma}
\end{eqnarray}
with $L_z \geq 2$.
This Hamiltonian, with $J_x=J_y=1$ for simplicity, may be solved for free
boundary conditions in $z$-direction by using the same variable transformation
as in the $1d$ Ising model.  When expressed in terms of the new product spin
variables $\tau$ the Hamiltonian for free boundary conditions can be seen to be
that of a stack of $2d$ Ising models with nearest-neighbour in-plane
interactions. The differences in the treatment of free and periodic boundary
conditions that are manifest in the $1d$ model also appear here, so  we treat
each separately.

\subsection{Free Boundary Conditions in $z$-Direction}

For free boundary conditions in $z$-direction (the case originally discussed by
Suzuki~\cite{suzuki_old}) we define bond spin variables $\tau_{x,y,z} =
\sigma_{x,y,z} \sigma_{x,y,z+1}$ on each vertical lattice bond in a cuboidal $L
\times L \times L_z$ lattice. The $\sigma$- and $\tau$-spins are  related by
\begin{equation}
\tau_{x,y,1} = \sigma_{x,y,1} \, \sigma_{x,y,2} \, , \, \ldots \, , \;
\tau_{x,y,L_z-1} = \sigma_{x,y,L_z-1} \, \sigma_{x,y,L_z}\, , \;
\tau_{x,y,L_z} = \sigma_{x,y,L_z}\; ,
\end{equation}
with an inverse relation of the form
\begin{equation} 
\sigma_{x,y,z} = \tau_{x,y,L_z} \, \tau_{x,y,L_z-1} \, \tau_{x,y,L_z-2} \cdots \tau_{x,y,z}\;,
\end{equation}
where  a one-to-one correspondence between the $\sigma$- and $\tau$-spin
configurations is maintained by specifying that the value of the $\sigma,
\tau$-spins on a given horizontal plane  (in this case $z=L_z$, i.e.,
$\tau_{x,y,L_z} = \sigma_{x,y,L_z}$) are equal. The resulting Hamiltonian is
missing one layer of spins,
\begin{equation}
H_{\rm fuki\mhyphen nuke} (\{ \tau \} ) = - \sum\limits_{x=1}^{L}\sum\limits_{y=1}^{L}\sum\limits_{z=1}^{L_z-1} \left( \tau_{x,y,z}  \tau_{x+1,y,z} + \tau_{x,y,z} \tau_{x,y+1,z} \right)\;,
\label{eq:stack1}
\end{equation}
so summing over these gives an additional factor of $2^{L \times L}$ in the
partition function (corresponding to the factor of $2$ in Eq.~(\ref{1DF})),
\begin{eqnarray}
    Z_{\rm fuki\mhyphen nuke} &=& \sum_{\{\tau\}} \exp\left(-\beta H_{\rm fuki\mhyphen nuke}(\{\tau\})\right)\nonumber\\
    &=& 2^{L^2}\sum_{\{\tau_{x,y,z\neq L_z}\}}\prod_{z=1}^{L_z-1}\exp\left(\beta \sum\limits_{x=1}^{L}\sum\limits_{y=1}^{L}\left( \tau_{x,y,z}  \tau_{x+1,y,z} + \tau_{x,y,z} \tau_{x,y+1,z} \right)\right)\nonumber\\
    &=& 2^{L^2}\prod_{z=1}^{L_z-1}\sum_{\{\tau_{x,y}\}_z}\exp\left(\beta \sum\limits_{x=1}^{L}\sum\limits_{y=1}^{L}\left( \tau_{x,y,z}  \tau_{x+1,y,z} + \tau_{x,y,z} \tau_{x,y+1,z} \right)\right)\nonumber\\
    &=& 2^{L^2}\prod_{z=1}^{L_z-1} Z _{2d\;\rm Ising} = 2^{L^2} \left(Z_{2d\;\rm Ising}\right)^{L_z-1},
    \label{eq:fukinuke-free}
\end{eqnarray}
where $\{\tau_{x,y}\}_z$ denotes summation over all $\tau$-spins with a given
$z$-component and $Z_{2d\;\rm Ising}$ is the standard partition function of the
$2d$ Ising layer. The boundary conditions in $x$- and $y$-directions are
arbitrary, as long as boundaries of different layers are not coupled, i.e.,
boundary conditions have no dependence on $z$ (the explicit notation in
Eqs.~(\ref{eq:stack1})~and~(\ref{eq:fukinuke-free}) assumes periodic boundary
conditions, but other conditions would carry through the calculation, too).
By taking the limit of infinite layers (but keeping $L_z$ fixed), one easily
arrives at
\begin{equation}
	\beta f_{\rm fuki\mhyphen nuke} \equiv -\lim_{L\rightarrow\infty}\frac{1}{L^2 L_z} \ln Z_{\rm fuki\mhyphen nuke} 
= \beta f_{2d\;\rm Ising} -\frac{\ln 2 + \beta f_{2d\;\rm Ising}}{L_z}\;,
\end{equation}
displaying explicitly the free-energy contributions of the two free surfaces at
$z=1$ and $z=L_z$ in terms of the (reduced) free-energy density $\beta
f_{2d\;\rm Ising} \equiv -\lim_{L\rightarrow\infty} \frac{1}{L^2} \ln
Z_{2d\;\rm Ising}$ of the $2d$ Ising model.

\subsection{Periodic Boundary Conditions in $z$-Direction}

\begin{figure}[t]
\centering	
  \includegraphics[scale=0.9]{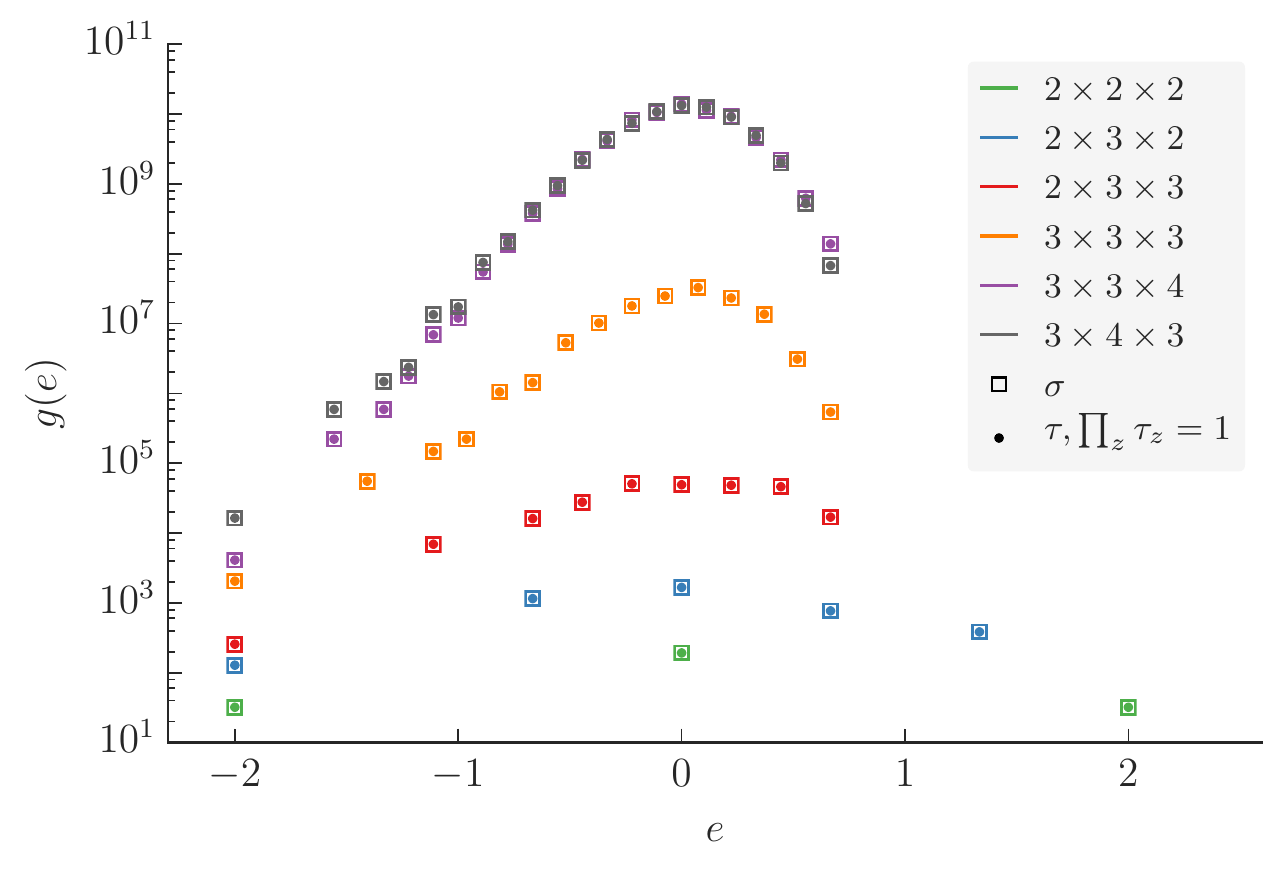}
  \caption{Number of states $g(e)$ over normalized energy
    \mbox{$e=E/(L_x\times L_y \times L_z)$} for the two representations of the
    fuki-nuke Hamiltonian with different lattice geometries under periodic
    boundary conditions. Boxes mark the number of states with a given energy
    $e$ for the $\sigma$-representation, dots mark the (rescaled) number of
    states $2^{L_xL_y}g_\tau(e)$ of states with energy $e$ in the
    $\tau$-representation. Since all dots fall into a box, the numbers agree.}
  \label{fig:sigma-tau-gE}
\end{figure}

We consider a cuboidal $L \times L \times L_z$ lattice with periodic
boundary conditions in $z$-direction, $\sigma_{x,y,L_z+1} = \sigma_{x,y,1}$.  We
define the bond spin variables $\tau_{x,y,z} = \sigma_{x,y,z} \sigma_{x,y,z+1}$
on each vertical lattice bond which must now satisfy the $L^2$ constraints 
$\prod_{z=1}^{L_z}\tau_{x,y,z} = 1$ because of the periodic boundary conditions.
The $\sigma$- and $\tau$-spins are subject to the inverse relation
\begin{equation} 
\sigma_{x,y,z} =\sigma_{x,y,1}\times \tau_{x,y,1} \, \tau_{x,y,2} \, \tau_{x,y,3} \cdots \tau_{x,y,z-1}\; .
\end{equation}
As for the $1d$ Ising model with periodic boundaries the $\sigma$-$\tau$
mapping is two-to-one. Since the transformation is carried out for each spin
lying in a horizontal $2d$ plane the $\tau$ partition function acquires an
additional factor of $2^{L\times L}$ arising from the transformation.  The
resulting Hamiltonian with $J_x = J_y = 1$ in terms of the $\tau$-spins is
again simply that of a stack of  $2d$ Ising layers with standard
nearest-neighbour in-layer interactions in the horizontal planes,
\begin{equation}
H_{\rm fuki\mhyphen nuke} (\{ \tau \} ) = - \sum\limits_{x=1}^{L}\sum\limits_{y=1}^{L}\sum\limits_{z=1}^{L_z} \left( \tau_{x,y,z}  \tau_{x+1,y,z} + \tau_{x,y,z} \tau_{x,y+1,z} \right)\;,
\label{eq:stack2}
\end{equation}
subject to  the $L^2$ constraints 
\begin{equation}
\prod_{z=1}^{L_z}\tau_{x,y,z} = 1,\qquad x=1,\ldots,L,\; y=1,\ldots,L\;.
\label{eq:fuki-nuke-tau-constraints}
\end{equation}
We collect numerical evidence in Fig.~\ref{fig:sigma-tau-gE}, that the variable
transformation is genuinely following the same pattern as in the $1d$ and $2d$
cases discussed earlier. For very small lattices we exactly enumerated the
models in Eqs.~(\ref{eq:fuki-nuke-sigma}) and~(\ref{eq:stack2}),
(\ref{eq:fuki-nuke-tau-constraints}) with the different spin representations
for periodic boundaries.  For some of the tested $3d$ lattice geometries with
dimensions $\left( L_x, L_y, L_z\right)$ with $L_i \leq 4$ we compare in
Fig.~\ref{fig:sigma-tau-gE} the number of states $g_{\sigma}(E)$ with an energy
$E = H(\{\sigma_i\})$. States that do not satisfy the $L_x\times L_y$
constraints in Eq.~(\ref{eq:fuki-nuke-tau-constraints}) are discarded during the enumeration to
yield the number of states $g_\tau(E)$ for the $\tau$-representation. Finally,
we respect the factors of $2$ from the transformation for the comparison,
$g_{\sigma}(E) = 2^{L_x L_y}g_\tau(E)$. For such small lattices, boundary
effects yield the most prominent contributions.  We also checked that our
program yielded the same results when  $L_x$ and $L_y$ were exchanged (not
shown).  We find that the (integer) numbers perfectly agree in all cases.

To interpret the role of the constraints we employ formally the same trick from the $1d$ Ising model of
rewriting the constraints in the partition function,
\begin{eqnarray}
	\label{eq:fuki-nuke-PBC}
    Z_{\rm fuki\mhyphen nuke} &=& 2^{L^2}\sum_{\{\tau\}} \exp\left(-\beta H_{\rm fuki\mhyphen nuke}(\{\tau\})\right)\prod_{x=1}^L\prod_{y=1}^L\delta \left( \prod_{z=1}^{L_z} \tau_{x,y,z}, 1 \right)
    \nonumber\\
    &=& \sum_{\{\tau\}} \exp\left(-\beta H_{\rm fuki\mhyphen nuke}(\{\tau\})\right)\prod_{x=1}^L\prod_{y=1}^L\left( 1 + \prod_{z=1}^{L_z} \tau_{x,y,z} \right)  .
\end{eqnarray}
If we expand the $\prod_{x=1}^L\prod_{y=1}^L\left( 1 + \prod_{z=1}^{L_z}
\tau_{x,y,z} \right)$ term in Eq.~(\ref{eq:fuki-nuke-PBC}) with the common
definition of the expectation value $\langle O \rangle_Z = Z^{-1}
\sum_{\{\tau\}} O e^{-\beta H}$ of an observable $O$ with respect to the
Hamiltonian $H$ and partition function $Z = \sum_{\{\tau\}}e^{-\beta H}$,
we find
\begin{eqnarray}
Z_{\rm fuki\mhyphen nuke} 
  &=& \sum_{\{\tau\}}\exp\left(-\beta H_{\rm fuki\mhyphen nuke}(\{\tau\})\right)
      \left(1 + \sum_{x=1}^{L}\sum_{y=1}^{L}\prod_{z=1}^{L_z}\tau_{x,y,z} + 
      \mathcal{O}\left(\tau\tau\right)\right)\nonumber\\
      &=& Z_{\rm fuki\mhyphen nuke}^{*}
      \left(
      1 + \sum_{x=1}^{L}\sum_{y=1}^{L}\langle\prod_{z=1}^{L_z}\tau_{x,y,z}\rangle_{Z_{\rm fuki\mhyphen nuke}^{*}} +
      \mathcal{O}(\tau\tau) \right)
      ,
  \label{eq:Zfukipbccorrelations1}
\end{eqnarray}
where $Z_{\rm fuki\mhyphen nuke}^{*} = Z_{\rm fuki\mhyphen nuke,\;
free}/2^{L^2} = \left(Z_{2d\;\rm Ising}\right)^{L_z}$, similar to the
calculation in Eq.~\ref{eq:fukinuke-free}, but without the outer sum from the
extra plane (and $L_z \rightarrow L_z+1$). 
Noticing that the product of $\tau$'s factorizes over the layers, leads to the
simplification
\begin{eqnarray}
Z_{\rm fuki\mhyphen nuke} 
&=& \left(
  Z_{2d\;\rm Ising}\right)^{L_z}\left( 
  1 + \sum_{x=1}^{L}\sum_{y=1}^{L} \left(\langle\tau_{x,y}\rangle_{Z_{2d\;\rm Ising}}\right)^{L_{z}} +
  \mathcal{O}(\tau\tau)
  \right) .
\label{eq:Zfukipbccorrelations2}
\end{eqnarray}
Finally, assuming translational invariance (i.e., periodic boundaries in each
$2d$ Ising layer) the leading correction further simplifies to
\begin{eqnarray}
\label{eq:Savvidy-Jonsson-eigen}
  Z_{\rm fuki\mhyphen nuke} &=& \left(Z_{2d,\;{\rm Ising}}\right)^{L_z}\left(1 +
  L^2 C_1^{L_z} + \mathcal{O}\left(\tau\tau\right)\right) ,
  \label{eq:Zfukipbccorrelations}
\end{eqnarray}
with $C_1 = \langle\tau_{1,1}\rangle_{Z_{2d,\;{\rm Ising}}}$ being the
normalized one-point function, or magnetization of the $2d$ Ising model, with
its distinct features: it vanishes for finite lattices (layers), but due to
spontaneous symmetry breaking assumes a non-zero value in the low-temperature
phase when taking the thermodynamic limit in finite field prior to setting the
field to zero.  Here, in the fuki-nuke case, ``field'' corresponds in the
original formulation with spins $\sigma_{x,y,z}$ to the coupling constant of a
nearest-neighbour interaction in $z$-direction.
%

Similarly the $\mathcal{O}\left(\tau\tau\right)$ contribution in
Eqs.~(\ref{eq:Zfukipbccorrelations1})-(\ref{eq:Zfukipbccorrelations}) can
be written as
\begin{eqnarray}
\!\!\!\!\!\!	\mathcal{O}\left(\tau\tau\right) &=&	
	\frac{1}{2}\left(\sum_{x_1=1}^{L}\sum_{y_1=1}^{L}\sum_{x_2 = 1}^{L}\sum_{y_2 = 1}^{L} 
\langle\prod_{z=1}^{L_z}\tau_{x_1,y_1,z}\tau_{x_2,y_2,z}\rangle_{Z_{\rm fuki\mhyphen nuke}^{*}} - 1 \right) + \mathcal{O}\left(\tau\tau\tau\right)\nonumber \\
&=&\frac{1}{2}\left(\sum_{x_1=1}^{L}\sum_{y_1=1}^{L}\sum_{x_2 = 1}^{L}\sum_{y_2 = 1}^{L}
\left(\langle \tau_{x_1,y_1}\tau_{x_2,y_2} \rangle_{Z_{2d\;\rm Ising}}\right)^{L_z} - 1 \right) +
\mathcal{O}\left(\tau\tau\tau\right)
  \label{eq:fukiO2correction}
\end{eqnarray}
which is a sum over all two-point functions of the $2d$ Ising model and hence a
much more difficult expression to evaluate exactly \cite{mccoy1973}. Only the
next-neighbour correlation, being proportional to the internal energy, is
readily accessible for finite layers (with periodic boundary conditions) from
the Kaufman solution \cite{kaufman1949}.
Even if the power $L_z$ on each of the two-point functions in
Eq.~(\ref{eq:fukiO2correction}) would not be present, we would end up with the
expression for the (high-temperature) susceptibility of the $2d$ Ising model. A
closed-form expression for this is as yet unknown, although its properties have
been analysed carefully to high precision using series expansions of extremely
high order~\cite{isingsuscept}.  The next terms in
Eq.~(\ref{eq:fukiO2correction}) are of the form
\begin{eqnarray}
\!\!\!\!\!\!\!\!\left(\langle \tau_{x_1,y_1}\tau_{x_2,y_2}\tau_{x_3,y_3} \rangle_{Z_{2d\;\rm Ising}}\right)^{L_z}, \quad 
\left(\langle \tau_{x_1,y_1}\tau_{x_2,y_2}\tau_{x_3,y_3}\tau_{x_4,y_4}\rangle_{Z_{2d\;\rm Ising}}\right)^{L_z}, \quad \dots
\label{eq:fuki-higher-order}
\end{eqnarray}
for all possible combinations of $x_1,y_1,\dots,x_4,y_4,\dots$.

In summary, we have found that the products of vertical stacks of
$\tau_{x,y,z}$ spins in $\prod_{z=1}^{L_z} \tau_{x,y,z}$ arising from the
constraints due to periodic boundary conditions give contributions from (all)
$n$-point Ising spin correlation functions (with $n \le L^2$) in each layer to
$Z_{\rm fuki\mhyphen nuke}$.
While providing an explicit exact answer to the problem, this prevents the
straightforward calculation of a closed-form expression for the fuki-nuke model
with periodic boundaries in the manner of Eqs.~(\ref{dimers_periodic}) and
(\ref{dimers_periodic2}) for the case of the $2d$ plaquette model with periodic
boundaries.\footnote{For $L_z=2$, $Z_{\rm fuki\mhyphen
nuke}(\beta)=Z_{2d,\;{\rm Ising}}(2\beta)$, because spins on top of each other
must be equal to fulfil the constraints, giving twice the energy of the usual
$2d$ Ising system (as can be verified by the exact data in 
Fig.~\ref{fig:sigma-tau-gE}). In total this gives a rule to calculate the sum 
over all $n$-point correlation functions of the $2d$ Ising model by $Z_{2d,\;{\rm
Ising}}(2\beta)/\left(Z_{2d,\;{\rm Ising}}\left(\beta\right)\right)^2 - 1$.}

A similar representation for $Z_{\rm fuki\mhyphen nuke}$ for periodic boundary
conditions has been obtained previously by
Jonsson~and~Savvidy~\cite{SavvidyFuki-Nuke} in a purely geometrical
interpretation of the fuki-nuke model as a model for fluctuating random
(closed) surfaces \cite{savvidy1,savvidy2}. By developing a suitable loop
Fourier transformation they found the solution to the fuki-nuke partition
function from eigenvalues of the transfer matrix between loops in the different
layers (tracing the intersections with the closed surfaces).  These eigenvalues
can be expressed in terms of the partition function and correlation functions
of the $2d$ Ising model, which can be identified with the corrections appearing
in Eq.~(\ref{eq:Savvidy-Jonsson-eigen}).  The  exact finite-size solution with
periodic boundary conditions thus amounts to evaluating all $n$-point spin
correlation functions in the $2d$ Ising model.  This is a much more difficult
task~\cite{mccoy1973} than for the almost trivial case of free boundary
conditions in~Eq.~(\ref{eq:fukinuke-free}), where no such correlation functions
appear. It would be interesting to see how, in the latter case, such a
simplification might occur in the geometrical surface/loop picture, too.

\begin{figure}[t]
\centering	
  \includegraphics[scale=0.5]{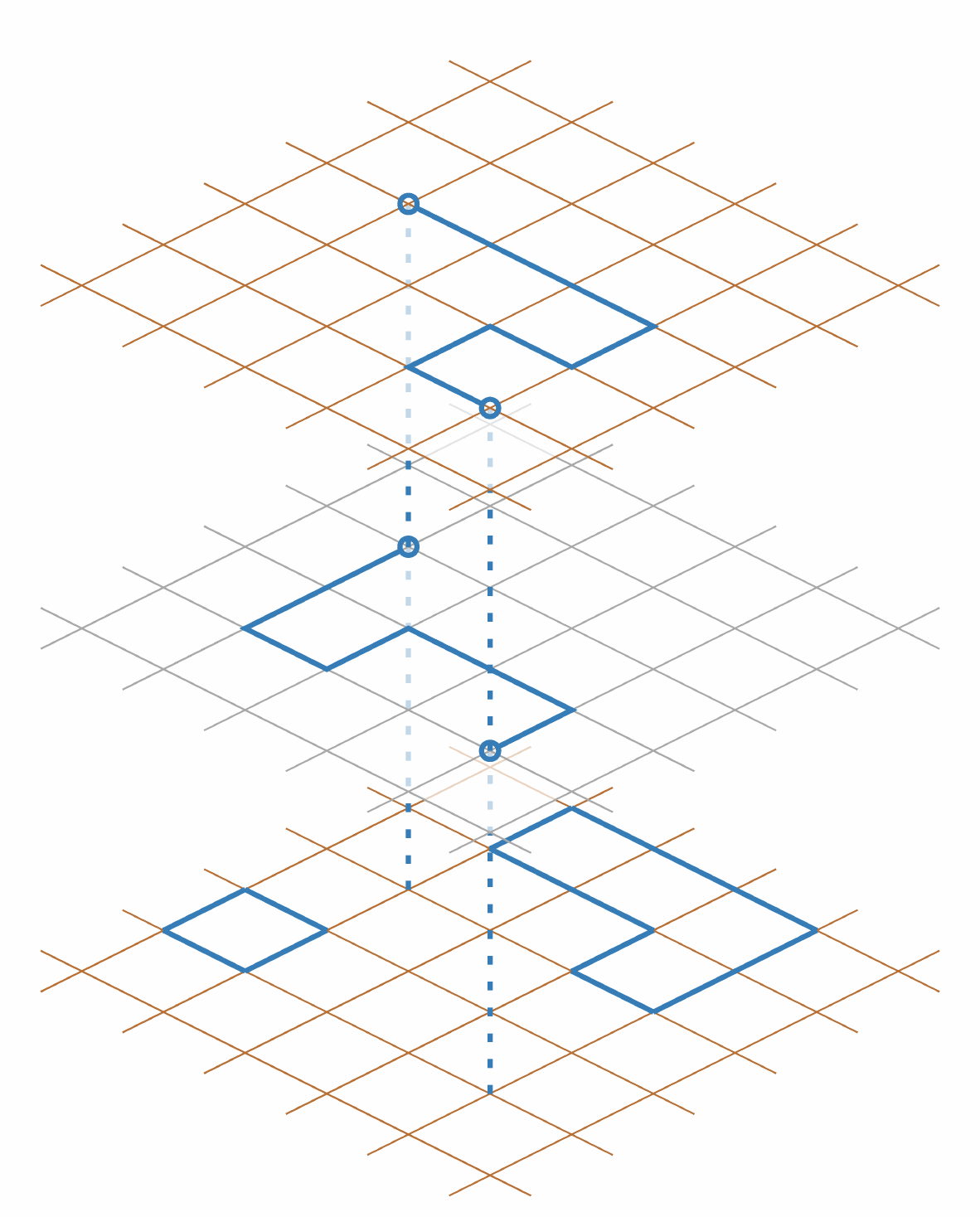}
  \caption{A dimer configuration of the fuki-nuke model with $L_z=3$ that can
    contribute to the partition function, although the mid and top layer have
    dangling ends (symbolised by open circles). These are connected through the
    constraints (dashed vertical lines) and contribute to the two-point function in 
    each of the two upper layers. Notice that additional,
    standard $2d$ Ising loops may appear, as those shown in the bottom layer,
    which are the standard contributions to the partition function of each
    layer.}
  \label{fig:dimers3d}
\end{figure}

The high-temperature expansion/dimer picture employed in Sec.~\ref{sec:dimergap} 
allowed an explicit solution of the $2d$ plaquette model with periodic
boundaries, where the constraints connect the different \emph{rows} of spins
with dangling ends (recall Fig.~\ref{Dimers}). We could employ a particle-gap 
symmetry there, easing the counting and effectively reducing the problem to a 
one-dimensional problem. A similar approach eludes us in $3d$ for the fuki-nuke 
model, however, where the equivalent picture leads to configurations with the 
constraints connecting the different \emph{layers}, see Fig.~\ref{fig:dimers3d}.  
The dimer configurations with two dangling ends in the mid and top layer
contribute to the two-point function.
Counting closed loops for the $2d$ Ising model is already a non-trivial 
combinatorial problem, and here we have to deal
with additional complexity depending on the number and position of the dangling
ends.  
It is obvious that the difficulty of the problem grows rapidly with the number
$n$ of dangling ends, contributing to the $n$-point function. While
Eqs.~(\ref{eq:Zfukipbccorrelations2})-(\ref{eq:fuki-higher-order}) give the
most explicit exact result, the high-temperature expansion/dimer approach is
the most intuitive pictorial way to explain how the constraints for periodic
boundary conditions induce the contributions of $n$-point correlations to the
partition function of each layer, as illustrated in Fig.~\ref{fig:dimers3d}.

That one set of boundary conditions should admit 
a closed-form
finite-size solution
and another not, is of course seen in other models, too. A canonical example is
the standard $2d$ Ising model where  the exact solution on finite lattices is
known only for  cases where there are (anti)periodic or twisted boundary
conditions in at least one direction \cite{kaufman1949,IsingBC}. For very
recent results on bulk, surface and corner free energies of the square lattice
Ising model for the case of free boundaries, see~\cite{baxter2016}.

\section{Conclusions}
\label{sec:conclusions}

Motivated originally by considerations from Monte Carlo simulations, where
periodic boundary conditions are  often employed in finite-size scaling studies
and where the density of states is of interest for multicanonical methods, we
investigated the differences between free and periodic boundary conditions in
calculating the partition function of various Ising models using  product spin
transformations. 

In $1d$ we observed that the partition function of the standard 
nearest-neighbour Ising model with periodic boundary conditions could be evaluated
using product spins if the constraint arising from the boundary conditions was
imposed via a convenient representation of the delta function. 

Similar considerations were found to apply to a $2d$ plaquette Ising model,
where the spin-bond transformations allowed exact evaluations of the
partition function for free and periodic boundary conditions. Although
equivalent to a $1d$ Ising model in the thermodynamic limit, the (boundary
condition dependent) finite-size corrections for the $2d$ plaquette model are
not identical.  

In $3d$ we compared the formulation of an anisotropic $3d$ plaquette model, the
fuki-nuke model, using product spin variables with free boundary
conditions~\cite{suzuki_old}  to the case of periodic boundary conditions
~\cite{castelnovo}. In understanding the detailed differences between these the
treatment of free and periodic boundary conditions  in the $1d$ Ising model and
$2d$ plaquette model provided a useful guide. For the fuki-nuke model the exact
finite-size partition function may be written as a product of $2d$ Ising
partition functions in the case of free boundary conditions using the product
variable transformation. A similar decoupling is not manifest with periodic
boundary conditions, where all $n$-point $2d$ Ising spin-spin correlations also
contribute to the expression for the $3d$ fuki-nuke partition function. 
As illustrated in Fig.~\ref{fig:dimers3d}, this can be most easily understood
in a pictorial way by employing the high-temperature expansion/dimer approach,
whereas the exact result in
Eqs.~(\ref{eq:Zfukipbccorrelations2})-(\ref{eq:fuki-higher-order}) displays the
contributing terms in the most explicit manner.  It is perhaps worth remarking
that the discussion of the fuki-nuke model in \cite{castelnovo} conflates the
discussion of free and periodic boundary conditions, although the overall
picture of a $2d$ Ising-like transition in the thermodynamic limit of the $3d$
fuki-nuke model remains, of course, correct in both cases.

The key point to be drawn from the various exact solutions explored in this
paper is that finite-size corrections due to  periodic boundary conditions may
be viewed as coming from induced correlations, which may be a useful point of
view when carrying out finite-size scaling analyses of numerical results.

\section*{Acknowledgements}
We would like to thank Nikolay Izmailian and Roman Koteck\'{y} for discussions
on, and providing references to,  various  boundary conditions for the $2d$
Ising model. Discussions with George Savvidy on the geometrical interpretation
of gonihedric models are appreciated and we would also like to  thank Lo\"ic
Turban for pointing out prior art in his lecture notes.  This work was
supported by the Deutsche Forschungsgemeinschaft (DFG) through the
Collaborative Research Centre SFB/TRR 102 (project B04), the
Deutsch-Franz\"osische Hochschule (DFH-UFA) through the Doctoral College
``$\mathbb{L}^4$'' under Grant No.\ CDFA-02-07, and by
the EU IRSES Network DIONICOS under Grant No.\ 612\,707.

\newpage
\appendix

\section{Two Dimensions: Helical Boundary Conditions by High-T Representation and Combinatorics}
\setcounter{figure}{0}

\label{helicalsolution}

Helical boundary conditions have already been used when comparing the $2d$
gonihedric Ising model with a $1d$ Ising model by means of  Metropolis Monte
Carlo simulations~\cite{davatolhagh}, although here the finite-size scaling
was not investigated since the focus was on the dynamical properties of the
model. 

We assume helical boundary conditions in $x$-direction, i.e., $\sigma_{L_x+1,y}
= \sigma_{1,y+1}$, and periodic boundaries in $y$-direction. The latter choice
is not arbitrary, because the next-to-nearest-neighbour interactions in the
Hamiltonian forbid helical boundaries in $y$-direction, or else one may find
different spins on the boundaries depending on whether one first goes along the
$x$-axis or $y$-axis. 

\begin{figure}[htb]
  \centering
    \subfigure[b][]{\raisebox{8mm}{
      \includegraphics[scale=0.7]{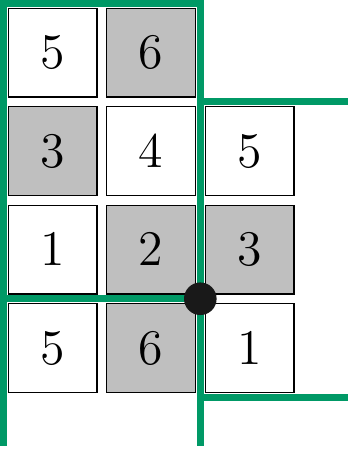}\hspace{4em}}}
    \subfigure[b][]{\raisebox{8mm}{
      \includegraphics[scale=0.7]{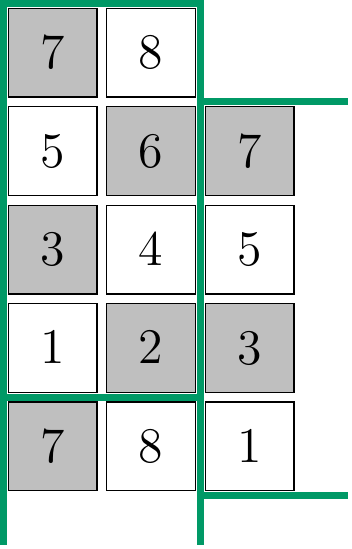}\hspace{4em}}}
    \subfigure[b][]{\raisebox{8mm}{
      \includegraphics[scale=0.7]{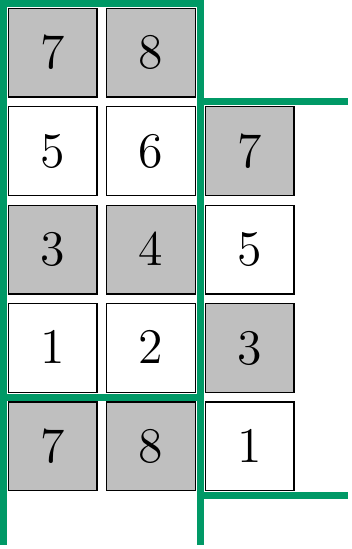}}}
    \caption{ Illustration of checker board configurations with helical boundaries
      along the $x$-direction and periodic boundaries in $y$-direction. The
      thick green lines separate repeating units of the system. Numbers
      distinguish the different plaquettes that are ``active'' (gray) or
      ``inactive'' (white).
      (a) For lattices with an odd number $L_y$ of plaquettes in $y$-direction,
      edges are created with spins that contribute to $3$ plaquettes (here, the
      black dot). Hence, that configuration does not appear
      in the partition function.
      (b) For even $L_y$, the checker board can be continued over the boundaries
      without having spins contribute with odd power.
      (c) In each column the gray and white plaquettes can be switched, leading 
      to another valid configuration. Here, the second column of (b) has been
      switched.
    }\label{fig:checkerboard}
\end{figure}

The partition function for helical boundaries can be found by counting the 
possible contributions when expanding the product in the high-temperature
representation in Eq.~(\ref{HighT}). As in the periodic case, only those
configurations can contribute to the partition function whose spins appear
with an even power. An arbitrarily chosen plaquette on an empty lattice has one
spin on each of the four corners and each spin contributes only once. For this
plaquette to contribute, adjacent plaquettes must also contribute, either
connected through a common bond or through a corner. Valid configurations are
thus either combinations of columns in $y$-direction that are closed through the 
periodic boundary conditions, one complete row that is closed with help of the 
helical boundaries or checker board configurations. Checker board configurations
only appear for lattices with an even number $L_y$ of spins in the direction of
the periodic boundaries, and here each column can have two possible patterns as 
depicted in Fig.~\ref{fig:checkerboard}. Hence, for odd $L_y$ we find
\begin{equation}
  Z_{2d,\;{\rm gonihedric},\;{\rm helical},\;{\rm periodic}} =
 2^{L_x L_y}\ch(\beta)^{L_x L_y}\left(1+\tnh(\beta)^{L_y}\right)^{L_x},
\end{equation}
and for lattices with even $L_y$,
\begin{flalign}
  \qquad 
  Z_{2d,\;{\rm gonihedric},\;{\rm helical},\;{\rm periodic}} &=\\
  2^{L_x L_y} \ch(\beta)^{L_x L_y}& \left(\left(1+\tnh(\beta)^{L_y}\right)^{L_x}
  + 2^{L_x}\tnh(\beta)^{L_x L_y/2}\right),\nonumber
\end{flalign}
where the additional term accounts for the contributions from checker-board-like
configurations, where the $L_x\times L_y/2$ plaquettes contribute a 
$\tnh(\beta)$ each. The freedom of column-wise switching of gray and white 
plaquettes is reflected in the prefactor $2^{L_x}$.

\section*{References}

\begin{thebibliography}{99}

	\bibitem{suzuki_old} M. Suzuki, Phys.~Rev.~Lett. {\bf 28} (1972) 507.    
	
	\bibitem{castelnovo} C.~Castelnovo, C.~Chamon and D.~Sherrington, Phys. Rev. B {\bf 81} (2010) 184303.
	
	\bibitem{suzuki1} Y. Hashizume and M. Suzuki, Int. J. Mod. Phys. B~{\bf 25} (2011) 73;\\
	Y. Hashizume and M. Suzuki, Int. J. Mod. Phys. B~{\bf 25} (2011) 3529.
	
	\bibitem{us_fukinuke} 
	D.~A. Johnston,
	J.~Phys.~A:~Math.~Theor.~{\bf 45} (2012) 405001;\\
	M.~Mueller, D.~A.~Johnston and W.~Janke,
	Nucl.~Phys.~B~{\bf 894} (2015) 1;\\
	D.~A.~Johnston, M.~Mueller and W.~Janke,
	Mod. Phys. Lett. B {\bf 29} (2015) 1550109.
	
	\bibitem{firstorder}
	D.~Espriu, M.~Baig, D.~A. Johnston and R.~P.~K.~C. Malmini,
	J.~Phys.~A:~Math.~Gen.~{\bf 30} (1997) 405.
	
	\bibitem{us_goni}
	M.~Mueller, W.~Janke and D.~A.~Johnston,
	Phys.~Rev.~Lett.~{\bf 112} (2014) 200601;\\
	M.~Mueller, D.~A.~Johnston and W.~Janke,
	Nucl.~Phys.~B~{\bf 888} (2014) 214.
	
	\bibitem{goni_glassy} 
	A.~Lipowski, J.~Phys.~A: Math.~Gen.~{\bf 30} (1997) 7365;\\
	A.~Lipowski and D.~A. Johnston, J.~Phys.~A: Math. Gen.~{\bf 33} (2000) 4451;\\
	A.~Lipowski and D.~A. Johnston, Phys.~Rev.~E~{\bf 61} (2000) 6375;\\
	M.~Swift, H.~Bokil, R.~Travasso and A.~Bray,  Phys.~Rev.~B~{\bf 62} (2000) 11494;\\
	D.~A. Johnston, A. Lipowski and R.~P.~K.~C. Malmini,
	in {\it Rugged Free Energy Landscapes: Common Computational Approaches to Spin Glasses, Structural Glasses and Biological Macromolecules}, ed.\ W.~Janke, Lecture Notes in Physics {\bf 736} (Springer, Berlin, 2008), p.~173.
	
	\bibitem{savvidy1}
	R.~V. Ambartzumian, G.~K. Savvidy, K.~G. Savvidy and G.~S. Sukiasian,
	Phys.~Lett.~B {\bf 275} (1992) 99;
	
	G.~K. Savvidy and K.~G. Savvidy,
	Mod.~Phys.~Lett.~A {\bf 08} (1993) 2963;
	
	G.~K. Savvidy and K.~G. Savvidy,
	Int.~J.~Mod.~Phys.~A {\bf 08} (1993) 3993.
	
	\bibitem{savvidy2}
	G.~K. Savvidy and F.~J. Wegner,
	Nucl.~Phys.~B {\bf 413} (1994) 605;
	
	G.~K. Savvidy and K.~G. Savvidy,
	Phys.~Lett.~B {\bf 324} (1994) 72;
	
	G.~K.~Bathas, E.~Floratos, G.~K.~Savvidy and K.~G.~Savvidy,
	Mod.~Phys.~Lett.~A {\bf 10} (1995) 2695;
	
	G.~K.~Savvidy, K.~G.~Savvidy and P.~G.~Savvidy,
	Phys.~Lett.~A {\bf 221} (1996) 233.

	\bibitem{loic} Lo\"ic Turban, in {\it Ph\'enom\`enes Critiques-V, Mod\`eles Exactement Solubles},\\ at {\tt http://gps.ijl.univ-lorraine.fr/webpro/turban.l}
	
	\bibitem{Ising} E.~Ising,  Z.~Phys. {\bf 31} (1925) 253.
	
	\bibitem{garrahan} R.~L.~Jack, L.~Berthier and J.~P.~Garrahan, Phys.~Rev.~E~{\bf 72} (2005) 016103.
  \bibitem{combinatorial2dIsing} M. Kac and J. C. Ward, Phys.~Rev.~{\bf 88} (1952) 1332;
    R. P. Feynman, \emph{Statistical Mechanics. A Set of Lectures} (The Benjamin and Cummings Publishing Co., Reading, Massachusetts, 1972).

  \bibitem{espriu_prats} D.~Espriu and A. Prats, Phys.~Rev.~E~{\bf 70} (2004) 046117.
  

  \bibitem{davatolhagh} 
  S.~Davatolhagh, D.~Dariush and L.~Separdar, Phys.~Rev.~E~{\bf 81} (2010) 031501.

  \bibitem{mccoy1973}
  B.~M.~McCoy and T.~T.~Wu, \emph{The Two-Dimensional Ising Model} (Harvard University Press, Cambridge, 1973);\\
  R.~J.~Baxter, \emph{Exactly Solved Models in Statistical Mechanics} (Academic Press, New York, 1982). 
		
  \bibitem{kaufman1949}
  B. Kaufman,  
  Phys. Rev. {\bf 76} (1949) 1232;\\
  A. E. Ferdinand and M. E. Fisher, Phys. Rev. {\bf 185} (1969) 832.

  \bibitem{isingsuscept}
  B. Nickel, J. Phys. A: Math. Gen. {\bf 32} (1999) 3889; \\
  B. Nickel, J. Phys. A: Math. Gen. {\bf 33} (2000) 1693; \\
  W. P. Orrick, B. G. Nickel, A. J. Guttmann and J. H. H. Perk, Phys. Rev. Lett. {\bf 86} (2001) 4120; \\
  W. P. Orrick, B. G. Nickel, A. J. Guttmann and J. H. H. Perk, J. Stat. Phys. {\bf 102} (2001) 795; \\
  S. Boukraa, A. J. Guttmann, S. Hassani, I. Jensen, J.-M. Maillard, B.  Nickel and N. Zenine, J. Phys. A: Math. Theor. {\bf 41} (2008) 455202;\\
  Y. Chan, A. J. Guttmann, B. G. Nickel and J. H. H. Perk, J. Stat. Phys. {\bf 145} (2011) 549.
	
  \bibitem{SavvidyFuki-Nuke}
    T. Jonsson and G. K. Savvidy, Phys.~Lett.~B {\bf 449} (1999) 253;\\ 
    T. Jonsson and G. K. Savvidy, Nucl.~Phys.~B {\bf 575} (2000) 661;\\ 
    G. K. Savvidy, J.~High~Energy~Phys. {\bf 09} (2000) 44;\\
    G. K. Savvidy, Mod.~Phys.~Lett.~B {\bf 29} (2015) 1550203.

	\bibitem{IsingBC}
	L.~Onsager, Phys.~Rev. {\bf 65} (1944) 117;\\
 	H.~J.~Brascamp and H.~Kunz, J.~Math.~Phys. {\bf 15}  (1974) 66;\\ 
	D.~L.~O'Brien, P.~A.~Pearce and S.~O.~Warnaar, Physica A {\bf 228}  (1996) 63;\\
	W.~T.~Lu and F.~Y.~Wu, Physica A {\bf 258} (1998) 157;\\
	W.~T.~Lu and F.~Y.~Wu, Phys.~Rev.~E {\bf 63} (2001) 026107;\\
	M.~C.~Wu and C.~K.~Hu, J.~Phys.~A: Math.~Gen. {\bf 35}  (2002) 5189;\\
	T.~M.~Liaw, M.~C.~Huang, Y.~L.~Chou, S.~C.~Lin and F.~Y.~Li, Phys.~Rev.~E {\bf 73} (2006) 055101(R);\\
  A.~Poghosyan, R.~Kenna and N.~Izmailian, Europhys. Lett. {\bf 111} (2015) 60010.

  \bibitem{baxter2016}
  R.~J.~Baxter, arXiv:1606.02029v3 [math-ph];\\
  E. Vernier and J. L. Jacobsen, J. Phys. A: Math. Theor. {\bf 45} (2012) 045003.

\end{thebibliography}
\end{document}